\newcommand\etal{et al. }
\newcommand\eg{e.g. }
\newcommand\ie{i.e. }
\newcommand\ein{{\it Einstein}}
\newcommand\ros{{\it ROSAT}}
\newcommand\xmm{{\it XMM-Newton}}
\newcommand\chandra{{\it Chandra}}
\newcommand\asca{{\it ASCA}}
\newcommand\sax{{\it BeppoSAX}}
\newcommand\bbxrt{{\it BBXRT}}
\newcommand\nh{\rm N_{H}}
\newcommand\xspec{{\small XSPEC}}

\documentclass{aa}
\begin{document}

   \title{X--ray emission line gas in the LINER galaxy M81
\thanks{Based on observations obtained with XMM-Newton, an ESA science 
    mission with instruments and contributions directly funded by 
    ESA Member States and the USA (NASA)}}
\authorrunning{M.J. Page \etal}

   \author{
M.J. Page\inst{1} 
\and
A.A. Breeveld\inst{1}
\and
R. Soria\inst{1}
\and
K. Wu\inst{1}
\and
G. Branduardi-Raymont\inst{1}
\and
K.O. Mason\inst{1}
%\and
%E.M. Puchnarewicz\inst{1}
\and
R.L.C. Starling\inst{1}
\and
S. Zane\inst{1}
          }

   \offprints{M.J.Page (mjp@mssl.ucl.ac.uk)}

   \institute{
$^{1}$ Mullard Space Science Laboratory, University College London,
Holmbury St Mary, Dorking, Surrey, RH5 6NT, UK
             }

   \date{Received 23-04-02; accepted 23-12-02}

\abstract{
We present the soft X--ray spectrum of the LINER galaxy M81 derived from a
long observation with the \xmm\ RGS. The spectrum is dominated by 
continuum emission
from the active nucleus, but also contains emission lines from
Fe L, and H-like and He-like N, O, and Ne. The
emission lines are significantly broader than the RGS point-source
spectral resolution; in the cross dispersion
direction the emission lines are detected adjacent to, as well as
coincident with, the active nucleus. This implies that they
originate in a region of a few arc-minutes spatial
extent (1 arc-minute $\sim$ 1 kpc in M81). The flux ratios of the OVII triplet
suggest that collisional processes are responsible for the line emission.
A good fit to the whole RGS spectrum is obtained using a model consisting of an
absorbed power law from the active nucleus and a 3 temperature optically thin thermal
plasma. Two of the thermal plasma components have temperatures of $0.18\pm
0.04$ keV and $0.64\pm 0.04$ keV, characteristic of the hot interstellar medium
produced by supernovae; the combined luminosity of the plasma at these two
temperatures accounts for all the unresolved bulge X--ray emission seen in the
\chandra\ observation by Tennant \etal (\cite{tennant01}). The third component
has a higher temperature ($1.7^{+2.1}_{-0.5}$ keV), and 
we argue that this, along with some of the 0.64 keV emission, comes
from X--ray binaries in the bulge of M81.
}

%
%________________________________________________________________

\maketitle

\section{Introduction}
\label{sec:introduction}

M81 is an Sab spiral galaxy hosting a low luminosity Seyfert nucleus
which shows the characteristics of a ``low ionization nuclear
emission-line region'' (LINER, Heckman \cite{heckman80}). LINERs make up a
significant fraction of all galaxies (between 1/5 and 1/3; Ho \etal \cite{ho97}), and because 
M81 is the nearest LINER it is an important target with which
to investigate the X--ray emission from such objects.

M81 has been the subject of a number of X--ray studies. It was first
observed with \ein\ (Elvis \& van Speybroeck \cite{elvis82}, Fabbiano
\cite{fabbiano88}) which resolved several discrete sources in M81, the
brightest of which was associated with the active nucleus. An
apparently diffuse emission component associated with the bulge of
M81, spatially extended over a few arc-minutes was detected first with
\ros\ (Roberts \& Warwick \cite{roberts00}, Immler \& Wang
\cite{immler01}) and confirmed using \chandra\ data 
(Tennant \etal \cite{tennant01}). Low resolution X--ray
spectroscopy of M81 from the \ein\ MPC and IPC suggested that the nuclear
X--ray source has an intrinsic column density of 
$> 10^{21} {\rm cm^{-2}}$ (Elvis \& van Speybroeck \cite{elvis82}, Fabbiano
\cite{fabbiano88}). Later, spectra from 
\ros\ (Immler \& Wang \cite{immler01}), \bbxrt\
(Petre \etal \cite{petre93}), \asca\ (Ishisaki \etal \cite{ishisaki96}) and
\sax\ (Pellegrini \etal \cite{pellegrini00}) indicated the presence of
a soft ($<$ 1 keV) thermal component in addition to an absorbed power law
component ($\Gamma \sim 2$) from the nucleus. 
In this paper we present the soft X--ray
spectrum of M81 at much higher resolution, from an observation with
the \xmm\ reflection grating spectrometer (RGS, den-Herder \etal
\cite{denherder01}).

\begin{figure*}
\begin{center}
\leavevmode
\setlength{\unitlength}{1cm}
\begin{picture}(13.8,13)
\put(0,0){\includegraphics{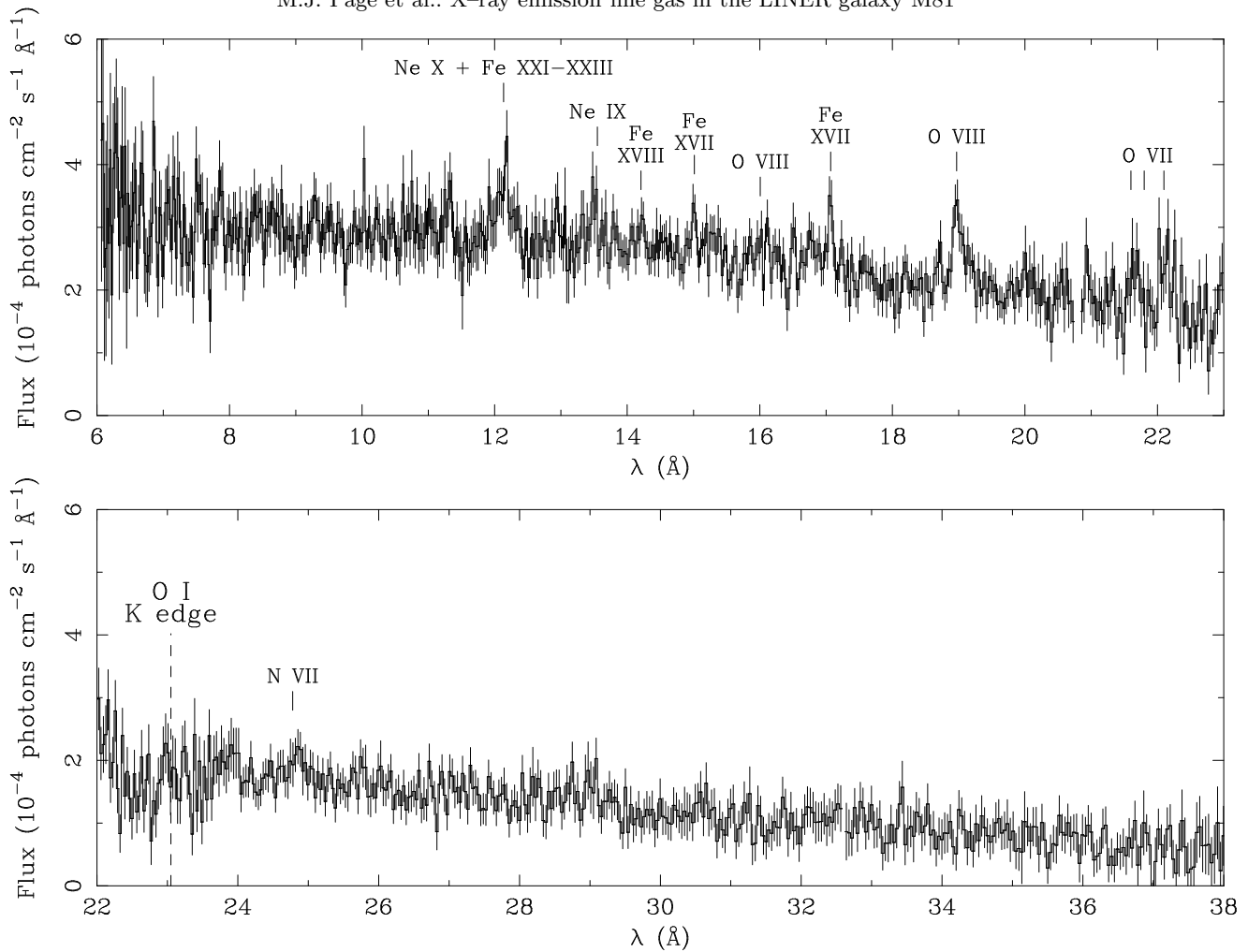}}
\end{picture}
\caption{The combined RGS spectrum with some prominent emission lines marked.}
\label{fig:rgsspectrum}
\end{center}
\end{figure*}

\section{RGS data reduction and spectral analysis}

\subsection{Data reduction}
\label{sec:observation}

M81 was observed by \xmm\ on the 22nd and 23rd April 2001 for a total of 
138 ks. 
The RGS data were reduced using the \xmm\ science analysis system (SAS)
public release version 5.2 and the latest calibration 
files (as of December 2001).
The source spectrum was
extracted from a region centred on the nucleus, enclosing 90\% of the point spread function in the
cross dispersion direction, 
while first and second order
selection was performed so as to include 93\% of the expected CCD pulse height
distribution of the source photons. Background regions were selected in the
cross-dispersion direction so as to exclude 99\% of the nuclear point spread
function; identical order selection was performed on nuclear and background
regions.  
Instrumental features such as that near the oxygen edge at $\sim$23 \AA\ were
corrected by dividing the effective area calibration in each response matrix 
by the model/data ratio from a power law 
fit to the RGS spectrum of the 
continuum source Mrk 421 (this correction is $<10\%$ over most of the
wavelength range). 
First and second order spectra from RGS1 and RGS2 and their corresponding
response matrices were resampled to match the RGS1 first order spectrum, 
combined, and rebinned by a factor of 3 (to $\sim 30$ m\AA\ bins) 
to improve signal to noise ratio before spectral analysis using \xspec.

\subsection{X-ray emission line gas}
\label{sec:emissionlines}

\begin{table}
\caption{Fluxes and equivalent widths of the most significant emission lines in
the RGS spectrum. The fluxes have been corrected for Galactic
absorption. Errors are quoted at 95\% for 1 interesting parameter.}
\label{tab:lines}
\begin{tabular}{llcc}
$\lambda$               & ion  & EW               & Flux\\
(\AA)                   &      & (m\AA)           & ($10^{-5}$ photon cm$^{-2}$ s$^{-1}$)\\
&&&\\
12.15$^{+0.04}_{-0.05}$ & Ne~X$^{*}$ & 61$^{+24}_{-26}$ & 2.0$^{+0.8}_{-0.8}$\\
13.49$^{+0.04}_{-0.04}$ & Ne~IX$^{**}$ & 34$^{+22}_{-24}$ & 1.1$^{+0.7}_{-0.8}$\\
14.22$^{+0.06}_{-0.09}$ & Fe~XVIII & 17$^{+13}_{-15}$ & 0.53$^{+0.41}_{-0.47}$\\
15.01$^{+0.03}_{-0.03}$ & Fe~XVII  & 27$^{+13}_{-18}$ & 0.83$^{+0.40}_{-0.56}$\\
17.06$^{+0.02}_{-0.02}$ & Fe~XVII  & 48$^{+20}_{-22}$ & 1.4$^{+0.6}_{-0.7}$\\
18.99$^{+0.04}_{-0.03}$ & O~VIII  & 118$^{+31}_{-32}$ & 2.5$^{+0.6}_{-0.7}$\\
21.67$^{+0.10}_{-0.10}$ & O~VII  & 51$^{+48}_{-45}$ & 1.3$^{+1.3}_{-1.2}$\\
22.12$^{+0.07}_{-0.07}$ & O~VII  & 79$^{+57}_{-55}$ & 2.0$^{+1.4}_{-1.4}$\\
24.86$^{+0.07}_{-0.10}$ & N~VII  & 49$^{+48}_{-34}$ & 1.1$^{+1.1}_{-0.8}$\\
&&&\\
\multicolumn{4}{l}{* blended with emission from Fe~XXIII}\\
\multicolumn{4}{l}{** blended with emission from Fe~XIX-XXI}\\
\end{tabular}
\end{table}

Fig. \ref{fig:rgsspectrum} shows the RGS spectrum. A number of prominent 
emission lines from H-like and He-like N, O, and Ne as well as L shell lines
from Fe XVII -- Fe XXIII are visible above the continuum; 
these are labelled in Fig. 
\ref{fig:rgsspectrum}, and the most significant lines are listed in
Table. \ref{tab:lines}. The lines are considerably broader
 than the RGS line spread function
for a
point source ($\sim$ 70 m\AA\ FWHM; den Herder \etal \cite{denherder01}). 
For example, O VIII Ly$\alpha$ at $\sim19$\AA\ is expected to 
have relatively little contamination from
neighbouring lines, but a gaussian fit to this line shows that its intrinsic 
width is 
inconsistent with $\sigma = 0$ at $> 99\%$.
This level of line broadening could occur if the gas is extended over a
region of $\sim$ 2 arc-minutes, because the RGS is a slitless dispersive
spectrograph and its resolution is degraded for extended
objects. Alternatively, the broadening could be due to kinematic motions in the
gas, for example if it is associated with the active nucleus.

\begin{figure}
\begin{center}
\leavevmode
\setlength{\unitlength}{1cm}
\begin{picture}(8.8,11.1)
\put(0,0){\includegraphics{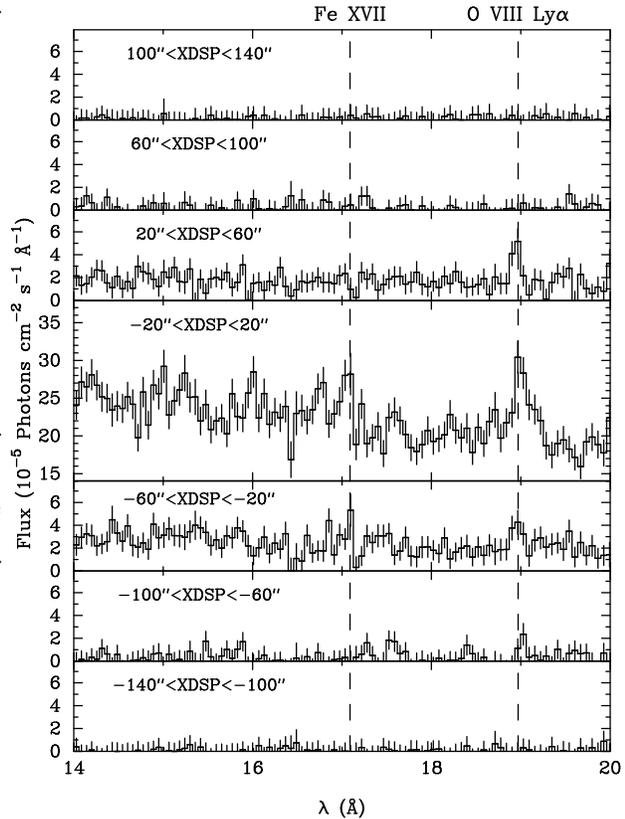}}
\end{picture}
\caption{Spectral cuts labeled with distance from the nucleus in the cross
dispersion direction (XDSP) demonstrating that the line emission comes from a
spatially extended region. The dashed lines correspond to prominent emission 
lines from FeXVII and OVIII.}
\label{fig:crosscuts}
\end{center}
\end{figure}

To distinguish between these alternatives, we have examined the spatial
distribution of the emission line gas in the RGS cross dispersion direction by
constructing spectra in 40 arc-second wide strips at a range of distances from
the nucleus of M81 (for comparison, the cross dispersion region used in Sect.
\ref{sec:observation} varies from $\sim 52$ arc-seconds wide around 20\AA\ to
$\sim 75$ arc-seconds wide near the ends of the spectrum). 
The spectra were binned to 60 m\AA\ and a background
spectrum was constructed from the average of the 
two most off-axis strips; this was subtracted from all the spectra. The
emission-line rich 14-20 \AA\ parts of these spectra are shown in
Fig. \ref{fig:crosscuts}. The emission line gas, and in particular O VIII
Ly$\alpha$, is extended over more than an arc-minute implying that the excess
broadening of the emission lines is due to their spatial extent, rather than
kinematic motions.

\begin{figure}
\begin{center}
\leavevmode
\setlength{\unitlength}{1cm}
\begin{picture}(8.8,10.1)
\put(0,0){\includegraphics{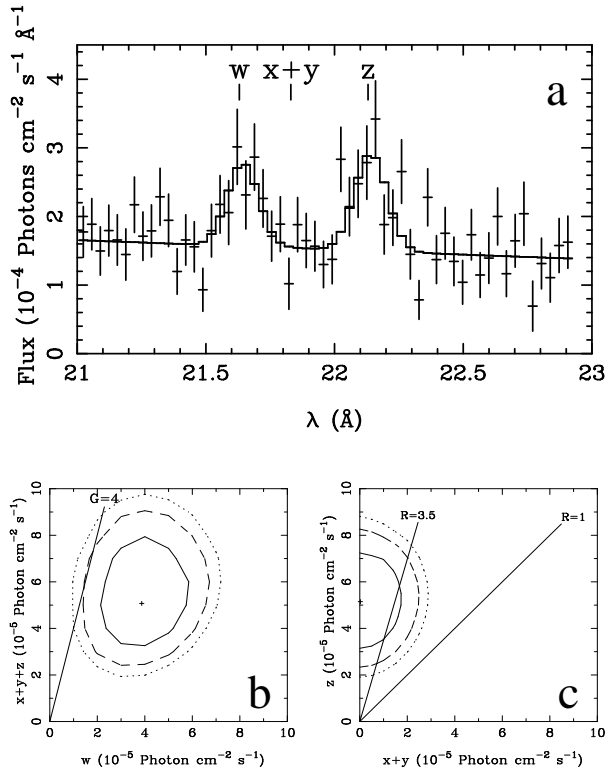}}
\end{picture}
\caption{
Panel a. shows the O VII triplet and best fit model.\ \ 
Panel b. shows the 68\% (solid),
90\% (dashed) and 95\% (dotted) confidence
contours for the intercombination (x+y) and forbidden (z) line strength
against the resonance (w) line strength; photoionization powered lines should
have line ratios to the left of the line G = (x+y+z)/w = 4.\ \  
Panel c. shows 68\% (solid),
90\% (dashed) and 95\% (dotted) confidence
contours for the forbidden (z) line strength against the intercombination (x+y)
line strength; the ratios R = z/(x+y) = 1 and R = 3.5 correspond to 
densities of
$\sim 10^{11} {\rm cm^{-3}}$ and $\sim 10^{9} {\rm cm^{-3}}$ respectively.}
\label{fig:oxygenplots}
\end{center}
\end{figure}

We further investigate the nature of the emission line gas using plasma
diagnositics from the He-like OVII triplet, as described in Porquet \& Dubau
(\cite{porquet00}). 
Taking the 21-23 \AA\ portion of the RGS spectrum, and assuming an underlying
power law continuum ($\Gamma=2$, see Sect. \ref{sec:modelling}), 
we fitted the 
OVII lines as 3 gaussians with the relative wavelengths of the resonance (w),
intercombination (x+y) and forbidden (z) lines fixed at their theoretical
ratios. The best fit has an acceptable $\chi^{2}/\nu = 56/51$, and is shown
superimposed on the 21-23 \AA\ RGS spectrum in Fig. \ref{fig:oxygenplots}a. 
In Fig. \ref{fig:oxygenplots}b we show the confidence contour of the (x+y+z)
against w line strength, a diagnostic of the ionization mechanism. 
The solid line shows the line ratio G 
= (x+y+z)/w = 4. As shown in figure 7 of Porquet \& Dubau 
(\cite{porquet00}), 
 plasmas in which photoionization is the dominant ionization mechanism are
expected to have ratios to the left of this line (\ie G $>$ 4); 
this is ruled out at 95\% confidence ($\Delta \chi^{2} > 4$ for one
interesting parameter G) in M81. 
For plasmas with line ratios to the right of this line, 
such as that observed in M81, collisional excitation
is important and
may be the dominant emission mechanism.
If the O~VII lines arise in a collisionally ionized plasma, the G ratio
provides a temperature diagnostic; by comparison with Porquet \etal 
(\cite{porquet01}) we find that the lines arise in a plasma with $kT <
0.26$~keV at 90\% confidence.

Fig. \ref{fig:oxygenplots}c 
shows confidence contours of the density diagnostic z
against (x+y). The line shown on the plot corresponding to ratio R
= z/(x+y) = 1 is excluded at $> 95\%$ along with all of parameter space to the
right of this line, implying (see Porquet \& Dubau
\cite{porquet00} figure 8) that the line emitting plasma has a density of $
{\rm n}_{e} <
10^{11} {\rm cm^{-3}}$. The best fit has very weak x+y lines relative to z, 
and consequently lies to the left of the line R = 3.5, hence the line ratio
observed in M81 is
best reproduced by a low density (${\rm n}_{e} < 10^{9} {\rm cm^{-3}}$) plasma.

\subsection{Modelling the whole RGS spectrum}
\label{sec:modelling}

It is evident from Fig. \ref{fig:rgsspectrum} that the emission lines in the
soft X--ray spectrum of M81 are superimposed on a strong continuum
source, almost certainly from the active nucleus at the heart of M81. 
Therefore we began by fitting a power law model with absorption 
(using the \xspec\ model TBabs of Wilms
\etal \cite{wilms00}) by the
line of sight Galactic column density $\nh=$ 4.16 $\times 10^{20}$
cm$^{-2}$ (Dickey \& Lockman \cite{dickey90}).  The parameters 
of this fit, and subsequent fits, are given in
Table \ref{tab:totalfits}.
We found that the power law model is a poor fit with 
$\chi^{2}/\nu = 1289/993$; 
the model overpredicts the data at both the high and low
wavelength ends of the spectrum, and has too small an O I edge at 23
\AA. We therefore added additional cold absorption, and obtained a much
better fit ($\chi^{2}/\nu = 1085/992$), with an additional $\nh\sim$ 5 $\times
10^{20}$ cm$^{-2}$.
This additional column density
confirms, and is
consistent with, previous results
(\eg Pellegrini \etal \cite{pellegrini00}) 
that the nucleus of M81 has some intrinsic 
absorption. 

We then attempted to model the emission line component of the spectrum. As
shown in Sect. \ref{sec:emissionlines} the O VII triplet is consistent with
an origin in a collisionally ionized gas, and the line emission comes from a
region extended over several kpc. We therefore began modelling the emission
lines by adding a single-temperature, solar-abundance `Mekal' thermal 
plasma model to the
absorbed power law model. This
resulted in a significant improvement in the fit ($\chi^{2}/\nu = 876/990$)  
for a
plasma temperature of $0.66\pm 0.04$ keV. 

\begin{figure*}
\begin{center}
\leavevmode
\setlength{\unitlength}{1cm}
\begin{picture}(8.8,9.8)
\put(0,0){\includegraphics{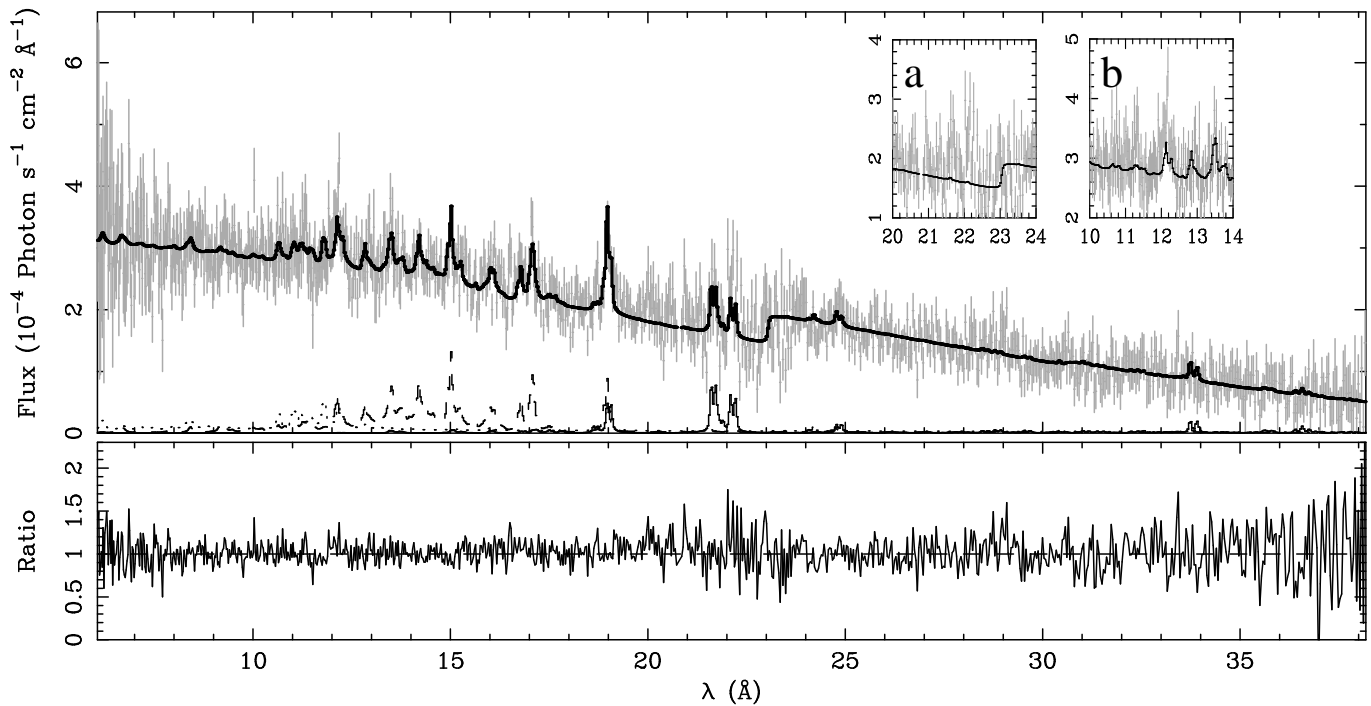}}
\end{picture}
\caption{Power law $\times$ 
intrinsic neutral absorption + 3 Mekal model fit to the
RGS spectrum. The top panel shows the data (grey points) and model (bold black line)
while the bottom panel shows the data/model ratio. 
Also shown in the top panel
is the contribution from the three different temperature thermal plasma
components: 0.18 keV (thin solid line), 0.64 keV (dashed line) and 1.7 keV
(dotted line). 
\ \ Inset a displays 
data and 
single-temperature Mekal model around the OVII triplet, showing that O VII 
lines are
not reproduced by the single temperature model.\ \ Inset b shows 
data and two-temperature Mekal model
around the Ne X / Fe XXI-XXIII blend, indicating 
that a third, higher-temperature
model component is required to reproduce these lines.}
\label{fig:rgsfit}
\end{center}
\end{figure*}

However, the emission lines are too
narrow in this model (the excessive width of the emission lines has already
been pointed out in Sect. \ref{sec:emissionlines}), and some emission lines,
particularly those from OVII, have much lower intensities in the 
 model than in the observed spectrum (see Fig. \ref{fig:rgsfit}). 
We addressed the first of these two 
problems by adding 2 extra Mekal
components, shifted by $\pm0.5\%$ in wavelength, with temperatures tied to
that of the first Mekal component, but with normalisations allowed to vary. 
This broadened the
 line profile of the model to mimic the degraded resolution 
of the RGS for the
extended emission line region.
%and resulted in a small, but significant reduction in
%$\chi^{2}$. 
We then added a second, lower
temperature, broadened Mekal component to the model. 
This resulted in another significant improvement in $\chi^{2}$,
reproducing most of the emission lines well, with the second plasma component
at a best fit temperature of $\sim 0.2$ keV, consistent with the temperature
deduced from the O~VII lines (Sect. \ref{sec:emissionlines}). 
However, although $\chi^{2}$ is
good, 
the blend of Ne X with Fe
XXI-XXIII at $\sim 12.1$ \AA\ is still underproduced by this model, and so a
third, higher temperature, broadened Mekal plasma was
added. This resulted in yet another significant improvement in the fit, to a 
final best $\chi^{2}/\nu = 803/980$, with a best fit temperature for the third 
Mekal component of $1.7^{+2.1}_{-0.5}$ keV. The model reproduces the spectrum
well, as shown in Fig.\ref{fig:rgsfit}.

We have not attempted to derive elemental abundances for the soft X-ray 
emission line gas, because the abundance ratios are somewhat degenerate with
the temperature structure of the gas. However, 
we have verified that our assumption of solar abundance is reasonable as 
follows. 
 Fe and O are responsible
for the strongest soft X--ray lines in M81, so the ratio of Fe to O will be the
most important abundance ratio in terms of the resultant emission line 
spectrum, and the easiest to determine. 
We have therefore repeated the three-Mekal fit, 
once with the Fe abundance halved, and once
with the Fe abundance doubled, but with the abundances of other
elements (including O) fixed at the solar values
 (Anders \& Grevesse \cite{anders89}).
 We find that with the Fe abundance halved, 
the best fit is worse by $\Delta \chi^{2} = 4$, while doubling the Fe abundance
results in a best fit that is worse by $\Delta \chi^{2} = 9$, with respect to
the three-Mekal, solar-abundance fit in Table \ref{tab:totalfits}. This
suggests that  the thermal plasma in M81 has an Fe to O 
ratio reasonably close to solar.

\begin{table}
\caption{Spectral fitting of the whole M81 RGS spectrum. All models include
 Galactic absorption of $\nh = 4.16\times 10^{20}$ cm$^{-2}$}
\label{tab:totalfits}
\begin{tabular}{lcccc}
(1)&(2)&(3)&(4)&(5)\\
Model&$\Gamma$ or $kT$&flux&$\nh$&$\chi^{2}/\nu$\\
&&&&\\
\hline
PL            &$1.67\pm0.02$&$105\pm 1$&&1289/993\\
&&&&\\
PL$\times\nh$        &$2.06\pm0.06$&$99.9\pm 3.8$&$4.8\pm0.8$&1085/992\\ 
&&&&\\
PL$\times\nh$     &$1.94\pm0.07$&$95.8\pm 3.7$&$3.3\pm0.8$&876/990\\ 
+Mek &$0.66\pm0.04$&$4.3^{+0.9}_{-0.7}$&&\\ 
&&&&\\
PL$\times\nh$     &$1.92\pm0.07$&$95.4\pm 3.7$&$3.1\pm0.8$&866/988\\ 
+Mek3 &$0.64\pm0.04$&4.7$^{+0.9}_{-0.7}$&&\\ 
&&&&\\
PL$\times\nh$    &$1.93\pm0.07$&$94.5\pm3.8$&$3.4\pm0.8$&821/984\\ 
+Mek3 &$0.20^{+0.04}_{-0.02}$&$1.5\pm 0.5$&&\\ 
+Mek3 &$0.66\pm0.05$&$4.2^{+0.8}_{-0.7}$&&\\ 
&&&&\\
PL$\times\nh$    &$1.94\pm0.06$&$91.6^{+7.4}_{-4.4}$&$3.4\pm0.8$&803/980\\ 
+Mek3 &$0.18\pm0.04$&$1.1^{+0.5}_{-0.4}$&&\\ 
+Mek3 &$0.64\pm0.04$&$4.2^{+0.8}_{-0.9}$&&\\ 
+Mek3 &$1.7^{+2.1}_{-0.5}$&$3.4^{+7.6}_{-2.2}$&&\\ 
&&&&\\
\hline
&&&&\\
\multicolumn{5}{l}{Explanation of columns:}\\
\multicolumn{5}{l}{1 PL = power law,}\\ 
\multicolumn{5}{l}{\ \ $\nh$ = neutral absorber (`tbabs'
model in \xspec),}\\
\multicolumn{5}{l}{\ \ Mek = thermal plasma (`Mekal' model in \xspec),}\\
\multicolumn{5}{l}{\ \  Mek3 = thermal plasma
(`Mekal' model in
\xspec) with}\\
\multicolumn{5}{l}{\ \  components
shifted by $\pm 0.5\%$ in wavelength to mimic the}\\
\multicolumn{5}{l}{\ \  broadening of the lines produced by the
spatial extent of}\\ 
\multicolumn{5}{l}{\ \ the emission region.}\\
\multicolumn{5}{l}{2 power law photon index or temperature of thermal}\\
\multicolumn{5}{l}{\ \ plasma (keV)}\\
\multicolumn{5}{l}{3 flux of model component in the energy range 0.3-2.0 keV,}\\
\multicolumn{5}{l}{\ \ in units of $10^{-13}$ erg cm$^{-2}$ s$^{-1}$
without Galactic}\\
\multicolumn{5}{l}{\ \ absorption 
of $\nh = 4.16\times 10^{20}$ cm$^{-2}$}\\
\multicolumn{5}{l}{4 intrinsic neutral column density ($10^{20}$ cm$^{-2}$).}\\
\end{tabular}
\end{table}

\section{Discussion}

In Sect. \ref{sec:emissionlines} we showed that the soft X--ray 
emission lines in M81 come from a region which is extended in both 
the dispersion direction (evidenced by the excess width
of the lines) and cross dispersion direction (see Fig.\ref{fig:rgsspectrum}) of
the RGS. We also showed that the diagnostic OVII triplet is emitted by a low
density plasma and is not powered by photoionization from the central AGN.
Furthermore, in the previous section we showed that 
a 3 temperature thermal plasma model provides
a good description of the emission line component of the M81 RGS spectrum in
conjunction with absorbed power law emission from the nucleus. 
 Two of the plasma components 
have well constrained temperatures of $0.18\pm0.04$ keV and $0.64\pm0.04$ keV, 
which are characteristic of the hot interstellar medium produced by 
supernova explosions; in this case we would expect these emission lines to 
come 
from a genuinely diffuse region (or regions) in the bulge of M81. 

X-ray imaging studies, first using \ros\ 
(Roberts \& Warwick \cite{roberts00}, Immler \& Wang
\cite{immler01}) and later using \chandra\ (Tennant \etal \cite{tennant01})
have indeed found apparently diffuse emission associated with the bulge of M81.
Furthermore, Immler \& Wang
(\cite{immler01}) extracted a PSPC spectrum of the M81 bulge using a 1 arc-minute 
radius
source circle, including both the active nuclues and the diffuse emission. They
found a good fit using an absorbed power law to represent the nucleus and a two
temperature  thermal plasma to 
represent the apparently diffuse emission. Their best fit plasma temperatures, 
$0.15\pm 0.02$ keV and $0.63\pm 0.11$ keV, are in extremely good
agreement with the temperatures obtained from the RGS spectrum. 
However
the spatial resolution of \ros\ ($\sim 5$ arc-second for the HRI and $\sim
20$ arc-second for the PSPC) 
was such that the emission from a large population of
weak point sources could easily be confused with genuinely diffuse emission. 

The much higher resolution observation performed with \chandra\ provides 
a much more
accurate determination of the truly diffuse emission from the nucleus. Tennant
\etal (\cite{tennant01}) found that of 
the bulge emission, even excluding the nucleus, 64\% comes from the resolved 
point
source population. The remaining unresolved bulge
emission reported by Tennant \etal (\cite{tennant01}) has a countrate of 0.092
s$^{-1}$ in the Chandra ACIS-S3 chip. 
The 0.18 keV and 0.64 keV thermal plasma components from our
best fit model in Table 1 would correspond to countrates of 0.036 s$^{-1}$ and 0.158
s$^{-1}$ in the same chip. The lowest temperature component alone is unable to
account for all the unresolved bulge emission, but the combinaton of the two
exceeds it.

This implies that some of the emission associated with the 0.64 keV thermal
plasma, and all of the emisson associated with the highest of the three 
plasma temperatures 
($1.7^{+2.1}_{-0.5}$ keV) must arise in part of the bulge population resolved 
by \chandra. We therefore propose low-mass X--ray binaries as the most likely 
origin 
of this higher temperature 
emission.

The line-of-sight velocity dispersion in the bulge of M81 is 
$165\pm10$~km~s$^{-1}$ (Prichet \cite{pritchet78}, Pellet \& Simien
\cite{pellet82}), so the 0.2 keV plasma has similar energy per unit mass
to the stellar component and is bound to the bulge. Using the
cooling curves from Landini \& Monsignori Fossi (\cite{landini90}) we find that
the cooling time for the 0.2 keV component is $\sim 4 \times 10^{7} / \sqrt{f}$
years where $f$ is the filling factor of the gas. This is much shorter than the
time since the last M81--M82
perigalactic passage ($\sim 5\times 10^{8}$ years ago, Brouillet
\etal \cite{brouillet91}; Chandar \etal \cite{chandar01}) and
hence the 0.2 keV gas, which has a mass of $10^{6}/\sqrt{f}\  M_{\sun}$, 
cannot be a relic of a nuclear starburst induced by the
passage of M82
because the gas must have been replenished or reheated much more recently.
The 0.64 keV component has a mass of $1.4\times 10^{6}/\sqrt{f}\ M_{\sun}$, and
potentially has a much longer cooling time, $\sim 3
\times 10^{5} / \sqrt{f}$ years, but it is too hot to be bound by the virial 
mass of the bulge. The sound crossing time for the bulge of
M81 is less than $10^{7}$ years, so the 0.64 keV gas must also be
replenished or reheated on a timescale of a few $\times 10^{7}$ years.

The simulations of Shelton
(\cite{shelton98}) allow us to estimate 
the approximate supernova rate that would sustain the diffuse 
X-ray luminosity of the bulge of M81. The combined flux from our two lowest
temperature thermal components, compared to the predicted \ros\ countrates from
section 7.4 of Shelton
(\cite{shelton98}) corresponds to supernova rates of between 
$\sim 4\times 10^{-3}$ year$^{-1}$ and $\sim 4\times 10^{-2}$ year$^{-1}$, 
where the
 lower estimate comes from the predicted countrate in the
lower energy R1 \ros\ band and the higher estimate comes from the predicted countrate in the higher energy R4 
\ros\ band. The flux in the \ros\ R4 band is dominated by the 0.64 keV plasma;
if we assume that half of this emission is due to resolved sources then the 
R4-band
estimate of the supernova rate is reduced 
to $\sim 2\times 10^{-2}$ year$^{-1}$. 
If the supernova rate in the bulge of
M81 is $\sim 4\times 10^{-3}$ year$^{-1}$ or higher we expect there to be
tens of supernova remnants within the bulge with ages $<
10^{4}$~years, although in the gas-poor environment of the bulge they may have
very low X-ray luminosities. There are 41 sources detected by {\it Chandra} within the bulge
of M81, of which $\sim 30\%$ have spectra which are softer than typical X-ray binaries
(Tennant \etal \cite{tennant01}). The collective flux of these sources is
sufficient to 
contribute significantly to the 0.64 keV plasma emission; 
further {\it Chandra} observations will be
required to determine whether these sources are persistent and thus potential supernova
remnants. 

UV imaging
carried out with the Hubble Space Telescope failed to reveal any individual
massive stars in the bulge (Devereux \etal \cite{devereux97}),
and this lack of a young, massive stellar population
leaves only type 1a supernovae as the possible origin of the X-ray emitting 
gas. We obtain the canonical
`expected' rate for type 1a supernovae in M81 from table 8 of Van den Bergh \& 
Tammann 
(\cite{vandenbergh91}). 
Assuming a distance of 3.6 Mpc, B magnitude of 7.31 after
correction for Galactic extinction (de Vaucouleurs \etal
\cite{devaucouleurs91}) and ${\rm H_{0}} = 75$ km s$^{-1}$ Mpc$^{-1}$, we
expect $\sim 6\times 10^{-3}$ type 1a supernovae year$^{-1}$ for the whole
galaxy, consistent with the rate of supernovae required to support the 
diffuse X-ray 
emission.

The diffuse X-ray emission may well relate to the optical line emission which
gives M81 its LINER classification.
The bulge of M81 is bright in H$\alpha$ emission (Devereux \etal 
\cite{devereux95}, Greenwalt \etal \cite{greenawalt98}), and in the absence of
 young massive stars Devereux \etal
(\cite{devereux97}) 
conclude that the diffuse H$\alpha$ emission must be powered
either by old post-AGB stars or by shocks, both of which are compatible with
the observed UV surface brightness. If the diffuse X-ray emission 
is indeed produced by supernova remnants, then fast shocks
must be propagating through the bulge. 

There are three pieces of
evidence that suggest shocks related to the X-ray emitting gas, rather than 
post-AGB stars, 
ionize the diffuse optical emission line 
gas.
Firstly, the H$\alpha$ emission region has a similar spatial extent to the 
X-ray line emitting
plasma, but it has an asymmetric, possibly spiral 
structure (Devereux \etal \cite{devereux95}, Greenwalt \etal 
\cite{greenawalt98}), and hence does not trace the underlying distribution of 
stars in the bulge, as might be expected if the gas is ionized by old post-AGB
stars.  
Secondly, the ionized H$\alpha$ emitting gas shows complex, non-rotational 
motion of
up to 200 km s$^{-1}$, as well as rotation at up to
300 km s$^{-1}$ (Goad \cite{goad76}). The non-rotational motion could be 
driven by the expansion of hot 
bubbles and suggests
mechanical heating of the gas. 
Thirdly, the optical emission line ratios suggest that photoionization by hot
stars is not
the ionization mechanism in the bulge. 
For example, the [S\,II]/(H$\alpha$+[N\,II])
ratio is much higher in the bulge of M81 than in any of the H\,II regions,
(Greenawalt \etal \cite{greenawalt98}), suggesting that photoionization by hot
stars is not
the ionization mechanism in the bulge. 
However shock heating, with shock
velocities of a few hundred km~s$^{-1}$, 
can result in large [S\,II]/H$\alpha$ ratios (Baum \etal
\cite{baum92}, Dopita \& Sutherland \cite{dopita95}) as observed. Furthermore,
the strength of [O\,III] $\lambda$5007 observed in the central few hundred pc 
of M81 cannot be reproduced by photoinization from hot stars (Golev \etal
\cite{golev96}; Wang \etal \cite{wang97}), 
but the [O\,III]/H$\beta$ ratio is similar to the 
predictions of the
 shock model of Dopita \& Sutherland (\cite{dopita95}) for shock velocities of
300~km~s$^{-1}$. 
In
this model, [O\,III] $\lambda$5007 is emitted by gas which is
photoionized by the shock-heated material. 
For a shock velocity of
300~km~s$^{-1}$ the shocked material has an electron temperature similar to 
the 0.2 keV plasma found in the RGS spectrum (Dopita \& Sutherland 
\cite{dopita96}). This suggests that the optical line-emission is related to
the soft X-ray line-emitting plasma, and that the ISM in the bulge of M81 is 
shock-heated, possibly by supernovae.

\section{Conclusions}

The \xmm\ RGS soft X-ray spectrum of M81 shows emission lines from H-like and
He-like N, O and Ne as well as Fe-L lines superimposed on a strong continuum
from the nucleus. The excessive width of the emission lines in the dispersion
direction, and their detection outside the nucleus in the cross dispersion
direction implies that the emission lines originate in a region of a few
arc-minutes spatial extent, corresponding to a few kpc in M81. The OVII triplet
line ratios suggest that photoionization is not the main ionization mechanism
and that collisional processes must be important in producing the observed
lines. The RGS spectrum can be fitted 
with a model consisting of an absorbed power
law from the nucleus and a 3 temperature optically thin thermal plasma. 
Two of the thermal
components have temperatures ($0.18\pm
0.04$ keV and $0.64\pm 0.04$ keV) 
which are 
consistent with the hot interstellar medium produced by supernovae; the
combined flux from these two components fully accounts for (in fact exceeds) 
the unresolved bulge
emission seen with \chandra. The X-ray emission could be produced by type 1a 
supernova
rates of $\sim 4 - 20 \times 10^{-3}$ year$^{-1}$, which is not unreasonable
for M81. We propose that the shocks generated by the supernova remnants could
also be responsible for the observed optical line emission in the bulge of 
M81. 
The third X-ray emitting thermal plasma component has a higher temperature
($1.7^{+2.1}_{-0.5}$ keV) and in order not to exceed the unresolved X-ray
emission found with \chandra\ we propose X--ray binaries in the bulge
of M81 as a likely origin of this emission, as well as 
$\sim$half of the 0.64 keV emission.


\begin{thebibliography}{}

\bibitem[1989]{anders89} Anders E. \& Grevesse N., 1989, Geochimica et 
Cosmochimica Acta, 53, 197

\bibitem[1992]{baum92} Baum S.A., Heckman T.M. 
\& van Breugel W., 1992, ApJ, 389,
208

\bibitem[1991]{brouillet91} Brouillet N., Baudry A., Combes F., Kaufman M. \&
Bash F., 1991, A\&A, 242, 35

\bibitem[2001]{chandar01} Chandar R., Tsvetanov Z. \& Ford H.C., 2001, ApJ, 122, 
1342

\bibitem[1995]{devereux95} Devereux N.A., Jacoby G. \& Ciardullo R., 1995, AJ,
110, 1115

\bibitem[1997]{devereux97} Devereux N., Ford H. \& Jacoby G., 1997, ApJ, 481, L71

\bibitem[1990]{dickey90} Dickey J.M. \& Lockman F.J., 1990, Ann. Rev.
    Ast. Astr., 28, 215

\bibitem[1995]{dopita95} Dopita M.A. \& Sutherland R.S., 1995, ApJ, 455, 468

\bibitem[1996]{dopita96} Dopita M.A. \& Sutherland R.S., 1996, ApJS, 102, 161

\bibitem[1982]{elvis82} Elvis M. \& van Speybroeck L., 1982, ApJ, 257, L51

\bibitem[1988]{fabbiano88} Fabbiano G., 1988, ApJ, 325, 544

\bibitem[1976]{goad76} Goad J.W., 1976, ApJS, 32, 89

\bibitem[1998]{greenawalt98} Greenawalt B., Walterbos R.A.M., Thilker D. \&
Hoopes C.G., 1998, ApJ, 506, 135

\bibitem[2001]{denherder01} den Herder J.W., et~al., 2001, A\&A, 365, L7

\bibitem[1996]{golev96} Golev V., Yankulova I., \& Bonev T., 1996, MNRAS, 
280, 29

\bibitem[1980]{heckman80} Heckman T.M., 1980, A\&A, 87, 152

\bibitem[1997]{ho97} Ho L.C., Filippenko A.V. \& 
Sargent W.L.W., 1997, ApJ, 487, 
568

\bibitem[2001]{immler01} Immler S. \& Wang Q.D., 2001, ApJ, 554, 202

\bibitem[1996]{ishisaki96}  Ishisaki Y., et~al., 1996, PASJ, 48, 237

\bibitem[1990]{landini90} Landini M., \& Monsignori Fossi B.C., 1990, 
A\&AS, 82, 229

\bibitem[2000]{pellegrini00} Pellegrini S., et~al., 2000, A\&A, 353, 447

\bibitem[1982]{pellet82} Pellet A., \& Simien F., 1982, A\&A, 106, 214

\bibitem[1993]{petre93} Petre R., Mushotzky R.F., Serlemitsos P.J., Jahoda K. 
\& Marshall F.E., 1993, ApJ, 418, 644

\bibitem[2000]{porquet00} Porquet D. \& Dubau J., 2000, A\&A Supp, 143, 495

\bibitem[2001]{porquet01} Porquet D., Mewe R., Dubau J., Raassen A.J.J. \&
Kaastra J.S., 2001, A\&A, 376, 1113

\bibitem[1978]{pritchet78} Pritchet C., 1978, ApJ, 221, 507

\bibitem[2000]{roberts00} Roberts T.P. \& Warwick R.S., 2000, MNRAS 315, 98

\bibitem[1998]{shelton98} Shelton R.L., 1998, ApJ, 504, 785

\bibitem[2001]{tennant01} Tennant A.F., Wu K., Ghosh K.K., Kolodziejczak J.J. 
\& Schwartz D.A., 2001, ApJ, 549, L43

\bibitem[1991]{vandenbergh91} van den Bergh S. \& Tammann G.A., 1991,
Annu. Rev. Astron. Astrophys., 29, 363

\bibitem[1991]{devaucouleurs91} de Vaucouleurs G., et~al., 1991, 
`Third Reference Catalogue of Bright Galaxies' Springer-Verlag

\bibitem[1997]{wang97} Wang J., Heckman T.M. \& Lehnert M.D., 1997, ApJ, 491, 114

\bibitem[2000]{wilms00} Wilms J., Allen A. \& McCray R., 2000, ApJ, 542, 914

\end{thebibliography}
\end{document}